\documentclass[10pt]{iopart}

\usepackage{graphicx}
\expandafter\let\csname equation*\endcsname\relax
\expandafter\let\csname endequation*\endcsname\relax
\usepackage{amsmath}
\usepackage{cite}
\usepackage{siunitx}
\usepackage{soul}
\usepackage{xcolor}
\begin{document}

\title[J. Opt. 000000]{Optically controllable coupling between edge and topological interface modes of graphene metasurfaces}

\author{Yupei Wang and Nicolae~C.~Panoiu}

\address{Department of Electronic and Electrical Engineering, University College London, Torrington Place, London WC1E 7JE, UK}
\ead{n.panoiu@ucl.ac.uk}
\vspace{10pt}
\begin{indented}
\item[]{\today}
\end{indented}

\begin{abstract}
Nonlinear topological photonics has been attracting increasing research interest, as it provides an exciting photonic platform that combines the advantages of active all-opticall control offered by nonlinear optics with the unique features of topological photonic systems, such as topologically-protected defect-immune light propagation. In this paper, we demonstrate that topological interface modes and trivial edge modes of a specially designed graphene metasurface can be coupled in a tunable and optically controllable manner, thus providing an efficient approach to transfer optical power to topologically protected states. This is achieved in a pump-signal configuration, in which an optical pump propagating in a bulk mode of the metasurface is employed to tune the band structure of the photonic system and, consequently, the coupling coefficient and wave-vector mismatch between edge and topological interface modes. This tunable coupling mechanism is particularly efficient due to the large Kerr coefficient of graphene. Importantly, we demonstrate that the required pump power can be significantly reduced if the optical device is operated in the slow-light regime. We perform our analysis using both \textit{ab initio} full-wave simulations and a coupled-mode theory that captures the main physics of this active coupler and observe a good agreement between the two approaches. This work may lead to the design of active topological photonic devices with new or improved functionality.
\end{abstract}

%
% Uncomment for keywords
%\vspace{2pc}
%\noindent{\it Keywords}: XXXXXX, YYYYYYYY, ZZZZZZZZZ
%
% Uncomment for Submitted to journal title message
%\submitto{\JPA}
%
% Uncomment if a separate title page is required
%\maketitle
%
% For two-column output uncomment the next line and choose [10pt] rather than [12pt] in the \documentclass declaration
\ioptwocol

\section{\label{sec:level1}Introduction}
The discovery and impact of quantum Hall effect and other phenomena of topological nature in condensed matter physics has spurred the extension of these ideas to photonics \cite{4,8}, resulting in the emergence of topological photonics \cite{7,5,9,6}. As part of these developments in topological photonics, new phenomena and applications, such as unidirectional, topologically protected light propagation in which disorder-induced backscattering is suppressed \cite{4}, have been demonstrated and reviewed \cite{8,7,5,9,6}. By analogy with the time-reversal symmetry breaking feature of quantum Hall effect, which induces gapless edge states \cite{11,12}, early research efforts have focused on the generation of topologically-protected edge states in magneto-optical photonic crystals (PhCs) under an external magnetic field \cite{13,14,15,16}. However, magneto-optic photonic crystals only have strong response to an applied magnetic field at microwave frequencies, so that extending these phenomena to the optical domain required alternative solutions. To this end, in time-reversal invariant photonic systems, phenomena analog to quantum spin-Hall \cite{18,19,20,21} and quantum valley-Hall \cite{24,22,23,25} effects have been demonstrated upon breaking the spin-conservation and spatial-inversion symmetries, respectively. Topologically-protected edge modes have been explored not only in the context of fundamental science, but also with the aim to fulfil their potential for new device applications, such as robust light-transmission devices \cite{26,17}, optical signal processing devices \cite{27}, time-delay lines \cite{28,29}, dispersion-engineering devices \cite{zyl21jo}, and sensors \cite{32}.

Most of the research in topological photonic structures has primarily been focused on the investigation of the linear optical response of such photonic systems. However, key functionalities of active photonic devices, such as tunability, optical frequency generation, and sensing, can most effectively be implemented by employing the nonlinear optical response of the underlying materials \cite{33}.To this end, active topological photonic devices relying on nonlinear optical effects, including Kerr effect \cite{34}, second-harmonic generation (SHG) \cite{36,37}, third-harmonic generation (THG) \cite{37,35,63}, and four-wave mixing (FWM) \cite{38}, have been successfully demonstrated. In particular, frequency-mixing processes between phase-matched topological edge modes, including SHG \cite{36,37,55}, FWM \cite{38} and THG \cite{37,55,63}, have been recently studied in PhCs and two-dimensional (2D) materials. In addition, applications of these nonlinear optical effects to imaging of topological edge states \textit{via} THG \cite{35}, lattice edge solitons \cite{39,40}, traveling-wave amplifiers \cite{41}, topological sources of quantum light \cite{42}, and topological insulator lasers \cite{30,31} have been proposed theoretically and in some cases demonstrated experimentally. 

Unique properties of graphene-based topological systems \cite{63,65}, in conjunction with strong nonlinear optical interactions occurring in graphene, can be used to develop nonlinear photonic devices and systems, which can be used to optically control topologically-protected defect-immune light flow at the nanoscale. Key factors, such as large, tunable carrier densities \cite{64} and long intrinsic relaxation times up to the picosecond range \cite{25}, make graphene an ideal platform for topological plasmonics at high frequency, low loss, and large topological bandgaps. In particular, due to large optical near-field enhancement and extended life-time of plasmons in graphene metasurfaces, SHG, THG, FWM, and other nonlinear optical processes of plasmonic edge states can be achieved in graphene metasurfaces at ultralow optical power \cite{38}.

A research topic of considerable practical importance is the development of efficient schemes for coupling light to optical modes of topological nature. The fact that this is by no means a trivial challenge, can be understood from the following example. Thus, by employing photonic valley-Hall effects, one can design optical waveguiding structures supporting valley-momentum-locked modes localized at the interface between two domains characterized by different valley Chern numbers \cite{47}. To excite these topological modes, one usually places a point-like source near the interface between the two domains so that the interface mode is excited and subsequently measured. In practice, however, this approach can be very ineffective, chiefly for two reasons. First, in addition to the topological mode, the interface can support regular modes, too, and these regular modes are generally excited together with the topological mode. Second, using a point-like source to excite an optical mode can be a rather inefficient approach, as the spatial overlap between the source field and the optical mode can be rather limited.

This example illustrates the importance of the development of efficient excitation techniques that would allow one to couple optical power from external sources to topological modes of photonic structures. Moreover, if such coupling techniques would permit a certain degree of tunability their functionality would be further enhanced. With these goals in mind, adopting graphene as the material platform to use for the implementation of these ideas is a natural choice, as it possesses ultra-fast response time \cite{43,44} and strong optical nonlinearity \cite{45,46}, which are optimal features when seeking to achieve device operation at low optical power and ultrafast tunability.

In this article, we design and theoretically analyze an effective scheme to couple light into an interface topological mode of a graphene metasurface. We demonstrate that, by taking advantage of the large Kerr nonlinearity of graphene, the device can be operated using low optical power and in a tunable manner. This functionality is achieved by using a pump-probe configuration, in which a pump field propagating in a bulk mode of the metasurface is used to tune its photonic structure and thus control the nonlinear optical coupling between a trivial edge mode and an interface topological mode in which a signal beam propagates. Importantly, we show that the required pump power can be significantly reduced if the device is operated in the slow-light (SL) regime. The conclusions of our rigorous computational investigation of the coupling mechanisms are supported by a coupled-mode theory describing the dynamics of the amplitudes of the two coupled modes. Our work can facilitate and spur the development of new or improved active photonic devices that combine the advantages provided by topological photonics and nonlinear optics.

The paper is organized as follows. In Sec.~\ref{sec2}, we present the geometry and optical characteristics of the graphene plasmonic waveguide investigated in this work. Then, in Sec.~\ref{sec3}, we discuss the optical properties of topologically protected valley transport in the proposed graphene plasmonic waveguide. This is followed in Sec.~\ref{sec4} by a quantitative analysis of the device characteristics. In particular, we describe theoretically and computationally the optically controllable nonlinear coupling between edge and topological modes of the graphene metasurface and its dependence on the mode parameters, such as frequency, wave vector mismatch, and pump power. Finally, in Sec.~\ref{sec5}, we summarize the main conclusions of our work.

\section{\label{sec2}Configuration of the graphene metasurface}
\begin{figure}[!t]
	\centering
	\includegraphics[width=\columnwidth]{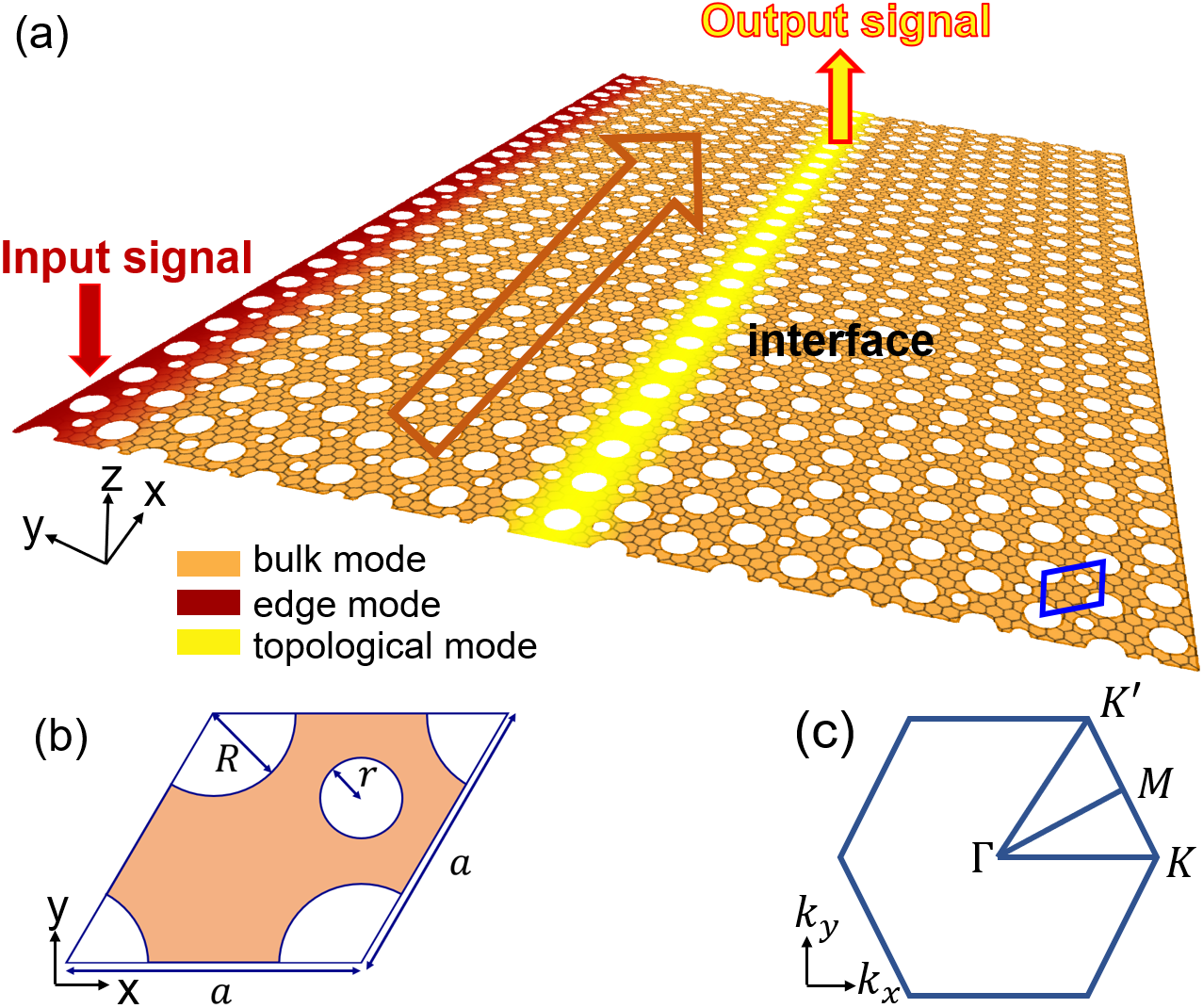}
\caption{(a) Schematic of the graphene metasurface plasmonic waveguide showing the coupling between the edge and topological interfacial modes. The graphene waveguide is created by adjoining together in a mirror-symmetric way two semi-infinite graphene metasurfaces consisting of a hexagonal periodic distribution of holes with primitive unit cell marked by blue rhombus. The red and yellow arrows indicate the optical signal coupled to the trivial edge mode (red) and the output signal collected from the topological mode (yellow), after it propagated along the domain-wall interface. The orange arrow indicates the pump beam propagating in the bulk mode, and used to control the optical coupling between the edge and interfacial modes. (b) Primitive unit cell of the infinite graphene metasurface with lattice constant, $a$, composed of two holes with radii $R$ and $r$. (c) First Brillouin zone of the graphene metasurface.}
	\label{cap1}
\end{figure}

The topological waveguide investigated in this work is implemented by using a composite graphene metasurface, as schematically illustrated in figure \ref{cap1}(a). The waveguide is created by adjoining together two semi-infinite graphene metasurfaces in a mirror symmetric manner, thus generating a domain-wall interface oriented along the $x$-axis, highlighted in yellow in figure \ref{cap1}(a). The two mirror-symmetric graphene metasurfaces consist of a periodic hexagonal array of holes in a uniform graphene sheet. In our analysis, we considered that the lattice constant was $a=\SI{500}{\nano\meter}$. The unit cell of the metasurface contains two holes, such that the spatial-inversion symmetry of the system is broken when the two holes have different radius, denoted here by $R$ and $r$. In this configuration, a plasmonic waveguide is formed at the interface between the two semi-infinite metasurfaces. As we will demonstrate later on, this waveguide can support both trivial and topological modes. Moreover, if the two metasurfaces are finite in the direction perpendicular onto the waveguide, trivial edge modes can exist at the boundaries of the two metasurfaces.

To control optically the frequency dispersion of the topological waveguide mode and its coupling to the edge modes, we use a pump beam that propagates in a bulk mode of the metasurface, illustrated by an orange arrow in figure \ref{cap1}(a). Specifically, the electrical permittivity of the metasurface can be controlled using the pump beam \textit{via} the Kerr effect in the graphene. The input red arrow in figure \ref{cap1}(a) indicates the excitation of the edge mode that carries the optical signal. Upon coupling to the topological waveguide mode, the signal propagates to the output of the waveguide where it is collected.

The optical dispersion relation of graphene is determined by its electric permittivity $\epsilon_g$, which can be evaluated from Kubo's formula \cite{46,48}:
\begin{eqnarray}\label{perm}
\epsilon_g(\omega)=&1-\frac{e^2}{4\epsilon_0\pi\hbar\omega h_g}\ln
{\frac{\xi-i\bar{\omega}}{\xi+i\bar{\omega}}}\nonumber\\
&+\frac{ie^2k_BT\tau}{\epsilon_0\pi\hbar^2\omega\bar{\omega}h_g}\left[
\frac{\mu_c}{k_BT}+2\ln\left(1+e^{-\frac{\mu_c}{k_BT}}\right)
\right]
\end{eqnarray}
In this equation, $T$ is the temperature, $\tau$ is the relaxation time, $h_g$ is the graphene thickness, $\mu_c$ is the chemical potential, whereas $\bar{\omega}=1-i\omega\tau$ and $\xi=2\tau|\mu_c|/\hbar$. In our simulations, we chose $T=\SI{300}{\kelvin}$, $\tau=\SI{50}{\pico\second}$, $h_g=\SI{0.5}{\nano\meter}$, and $\mu_c=\SI{0.2}{\electronvolt}$. The wavelength at which the graphene metasurface is operated was chosen to be in the THz range because at these wavelengths the linear and nonlinear optical losses of graphene, such as two-photon absorption (TPA), are negligible. Conceptually, our analysis remains qualitatively valid if the graphene metasurface is operated in a different spectral domain but, of course, the specific quantitative results will be different.

\section{\label{sec3}Band structure analysis and frequency dispersion properties of the waveguide modes}
\begin{figure}[t]
\centering
\includegraphics[width=\columnwidth]{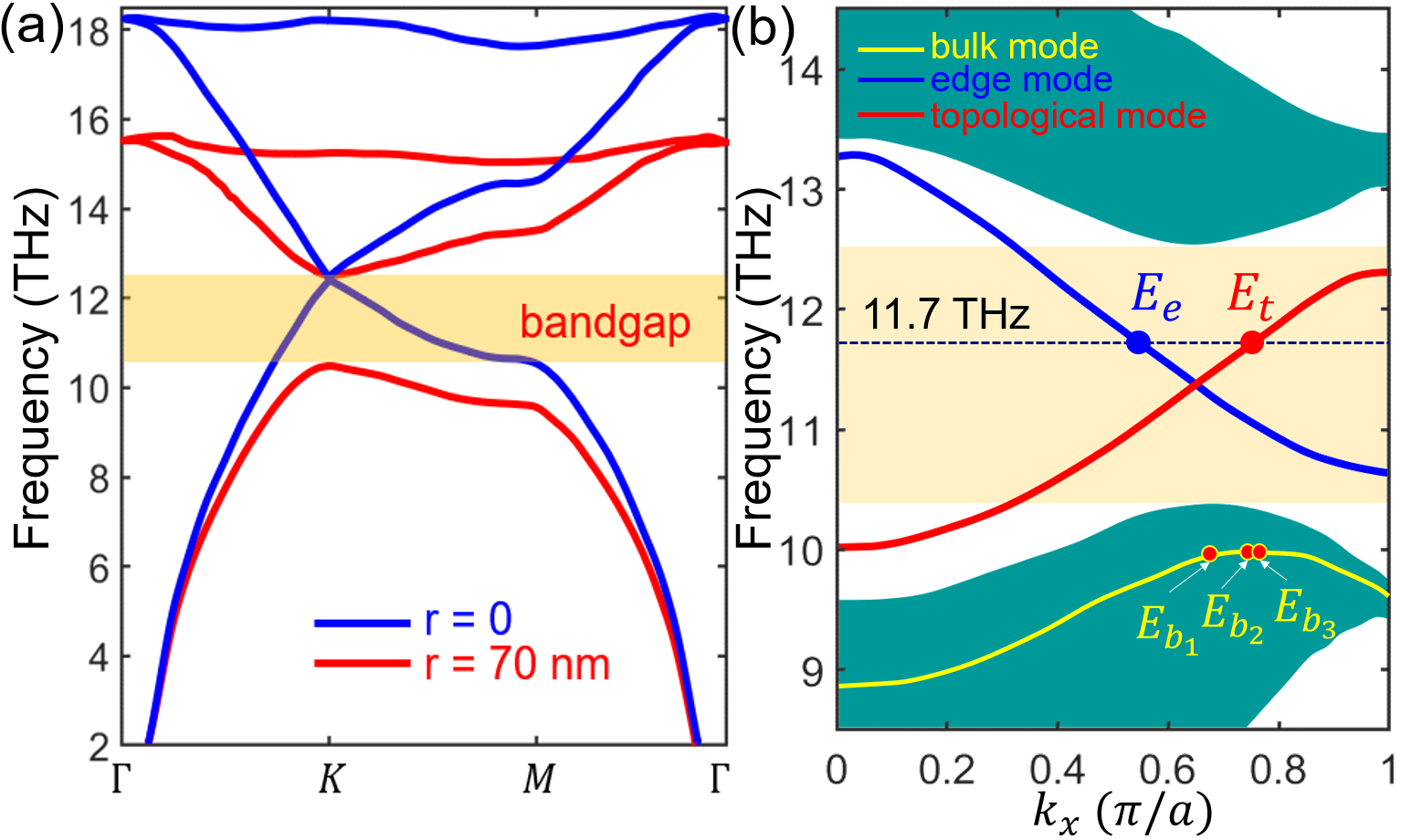}
\caption{(a) Band diagram of the graphene metasurface, determined for different radii of nanohole $r$. When $r=0$, a Dirac cone is formed at the crossing point of the first and second bands (blue lines). By introducing a second nanohole with $r=\SI{70}{\nano\meter}$ in the primitive cell of the metasurface, a nontrivial bandgap (yellow band) from \SIrange{10.4}{12.5}{\tera\hertz} is created. (b) Projected band structure of the composite metasurface, showing bulk bands (green) forming a continuum, a nontrivial topological mode (red) of the graphene plasmonic waveguide, and a trivial edge mode (blue). The smaller and larger holes have radius of $r=\SI{70}{\nano\meter}$ and $R=\SI{140}{\nano\meter}$, respectively. The frequency dispersion curve of a bulk mode, depicted in yellow, together with three points indicating photonic states with different group velocities that were used as pump modes, are also shown.}
\label{cap2}
\end{figure}

To demonstrate the existence of waveguide topological modes in our structured graphene metasurface, we have determined the photonic band structure of an infinite metasurface, as well as the projected band structure of the interface waveguide obtained by joining together in a mirror-symmetric manner two semi-infinite metasurfaces. These computations were performed with COMSOL Multiphysics\textsuperscript{\textregistered}, a commercially available software package \cite{COMSOL}. In the first step of our analysis we determined the linear optical response of the system by neglecting the optical nonlinearity of graphene. The photonic band diagram of an infinite metasurface, determined in the case when the metasurface has inversion symmetry, \textit{i.e.}, $R=\SI{140}{\nano\meter}$ and $r=0$, and for a noncentrosymmetric metasurface with $R=\SI{140}{\nano\meter}$ and $r=\SI{70}{\nano\meter}$, are plotted in figure \ref{cap2}(a) using blue and red curves, respectively. It should be noted that graphene metasurfaces with such geometrical parameters can be readily fabricated using widely available fabrication techniques, such as \textit{e}-beam lithography.

The calculated band diagrams show that in the case when the graphene metasurface has inversion symmetry ($r=0$, blue lines), that is when the lattice of holes belongs to the $C_{6v}$ point symmetry group, there exists a symmetry-protected Dirac cone at the $K$-symmetry point. To be more specific, the first and second bands cross each other forming a Dirac point with frequency of $\SI{12.5}{\tera\hertz}$, with the frequency dispersion nearby this Dirac point being linear. Moreover, the $C_{6v}$ point symmetry group is reduced to $C_{3v}$ symmetry group when holes with finite radius, $r\neq0$, are introduced in the unit cell of the graphene metasurface. One consequence of this change in symmetry is that a band gap opens at the frequency of the Dirac point. For example, for a radius of $r=\SI{70}{\nano\meter}$ [red lines in figure \ref{cap2}(a)], the frequency minimum of the second band remains practically unchanged whereas the frequency maximum of the first band decreases by about $\SI{2.1}{\tera\hertz}$. This bandgap, which opens upon breaking the inversion symmetry of the lattice, is topologically nontrivial \cite{25} and is indicated in figure \ref{cap2}(a) by the yellow strip. All these results remain qualitatively unchanged if the hole radius $r$ is varied, with the frequency bandgap becoming wider (narrower) when the radius $r$ increases (decreases). In our subsequent analysis, the radius of the smaller holes is fixed to $r=\SI{70}{\nano\meter}$.

The proposed graphene metasurface exhibits a nontrivial Berry curvature distribution in the momentum space around the two sets of nonequivalent high-symmetry valleys points $K$ and $K^{\prime}$ indicated in figure \ref{cap1}(c). In order to obtain a nonzero valley-dependent topological index, we can generate a finite difference of the Chern numbers ($+1$ or $-1$) between valleys at $K$ and $K^{\prime}$. This can be achieved by rotating half of the graphene metasurface by an angle of $\pi$ and placing it in a mirror-symmetric way near the other half of the metasurface, as per figure \ref{cap1}(a). As a result of this operation, the valleys at $K$ points of one metasurface are located directly opposite to the valleys at $K^{\prime}$ points of the other one, so that a finite difference of the valley Chern number across the domain-wall interface is achieved. As a consequence of this fact, valley-Hall topological modes can exist inside the nontrivial bandgap of this composite graphene metasurface.
\begin{figure}[!t]
\centering
\includegraphics[width=\columnwidth]{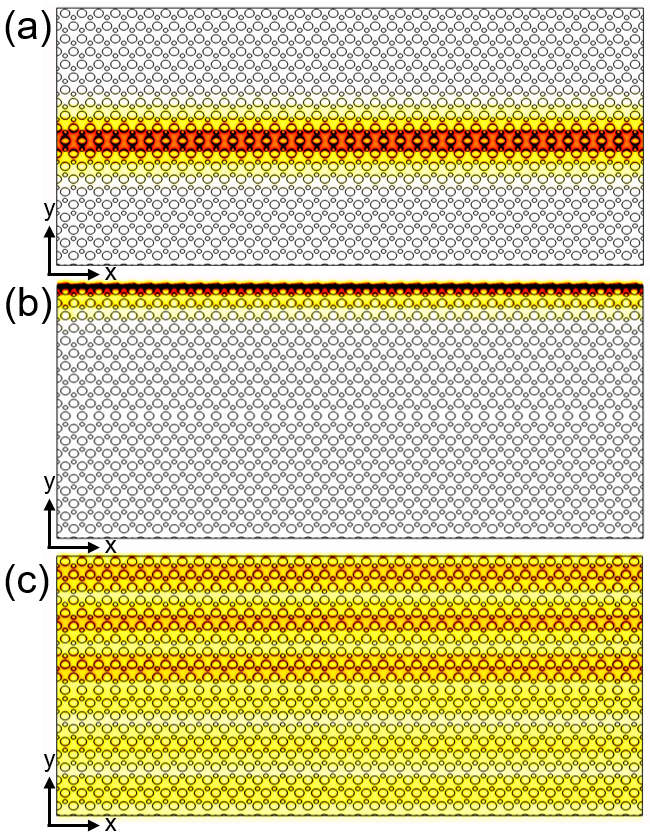}
\caption{From the top to bottom panel, field distribution of the topological interface mode corresponding to \SI{11.7}{\tera\hertz}, trivial edge mode determined at the same frequency [labelled $E_t$ and $E_e$ in figure \ref{cap2}(b), respectively], and the bulk mode labelled $E_{b_{1}}$ in figure \ref{cap2}(b), respectively.}
\label{cap3}
\end{figure}

The projected band structure of the composite metasurface that contains the 1D interfacial waveguide has been computed numerically and is presented in figure \ref{cap2}(b). In these calculations we used a supercell with a number of $20$ unit cells along the $y$-axis and periodic boundary conditions along $x$-axis. This means that the metasurface is finite along the transverse direction ($y$-axis) and infinite along the longitudinal one ($x$-axis). The specific value of the transverse width of the supercell was chosen such that the optical coupling between the topological and edge modes is strong enough to allow for efficient energy transfer between the two optical modes. Since the metasurface is periodic along the $x$-axis with period $a$, the wave vector component $k_x$ varies from $0$ to $\pi/a$. The green regions in figure \ref{cap2}(b) represent the projected bulk bands, with one such mode being indicated by the yellow curve, whereas the blue and red lines represent the linear dispersion curves of a trivial edge mode propagating at the edges of the metasurface and a topological valley waveguide mode located at the interface between the two metasurfaces, respectively. Also, the yellow band shows the nontrivial bandgap. Moreover, the dispersion curves of the trivial edge mode and waveguide topological mode cross each other, suggesting that, near the crossing frequency, optical power can be efficiently transferred between the two modes.

To further explore the optical properties of the modes of the finite, composite graphene metasurface (bulk, edge, and topological modes) we have determined their optical field profiles. In the case of the topological valley mode and trivial edge mode, we chose a frequency inside the bandgap, of $\SI{11.7}{\tera\hertz}$; the corresponding field distributions are depicted in figures \ref{cap3}(a) and \ref{cap3}(b), respectively. The optical field of the topological interfacial mode propagates along the waveguide formed by the two metasurfaces, whereas the optical field of the edge mode is located at the boundary of the metasurface. Note that the field distribution of both modes is highly localized, which is primarily due to the large electric permittivity of graphene. Moreover, we also plot in figure \ref{cap3}(c) the optical field profile of the bulk mode indicated in figure \ref{cap2}(b) by the yellow curve, determined at wave vector and frequency marked with the red dot labelled $``E_{b_{1}}"$. As expected, the optical field of the bulk mode spreads throughout the graphene metasurface, but the corresponding field intensity is markedly stronger in the upper half of the metasurface. Our simulation results show similar electric field profiles for the bulk mode determined at the other two points labelled by $``E_{b_{2}}"$ and $``E_{b_{3}}"$. However, despite the fact that the field profiles of the bulk mode do not change much among the points $E_{b_{1}}$, $E_{b_{2}}$, and $E_{b_{3}}$, the group velocity (GV) determined at these points has significantly different values . As it will become apparent later on, this has important consequences regarding the operation of our device, especially vis-\`{a}-vis its optical tunability.

\section{\label{sec4}Optically controllable nonlinear mode coupling \textit{via} Kerr effect}
The field distribution of the pump mode plotted in figure \ref{cap3}(c) suggests that the coupling between the edge and topological interfacial modes could be tuned by varying by dint of the Kerr effect the electrical permittivity of graphene in the region between the edge of the metasurface and the interface. Therefore, in this section we investigate the influence of the optical power propagating in the bulk pump mode on the dispersion characteristics of the edge and topological modes.

\subsection{Kerr tunability of the edge and topological waveguide modes}
Kerr effect is a nonlinear optical process ideal to use for tuning the optical properties of graphene, chiefly due to its extremely large second-order nonlinear refractive index, $n_2$. In particular, Kerr coefficient $n_2$ of graphene is more than 9 orders of magnitude larger than that of most optical materials customarily used in nonlinear optics \cite{49}. Additionally, other features of graphene, such as chemically and electrically tunable optical properties and strong nonlinear optical response in a broad frequency domain spanning from terahertz down to mid-infrared frequencies \cite{50,51}, are particularly attractive for nonlinear optics applications.

The change of the refractive index of graphene in response to an applied (pump) electric field $\mathbf{E}_p(\mathbf{r})$ is described by \cite{33}:
\begin{eqnarray}
\Delta n(\mathbf{r})=\frac{1}{2}c\epsilon_0 n n_2\left|\mathbf{E}_p(\mathbf{r})\right|^2,
\label{deln}
\end{eqnarray}
where $n$ is the index of refraction of graphene in the absence of the applied electric field and $n_2=\SI{6.3e-11}{\meter^2/\watt}$ is the second-order nonlinear refractive index of graphene \cite{52}. This formula is based on the assumption that TPA is not accounted for in our system, as we designed our device to operate at THz frequencies, namely in a spectral range where TPA in graphene is negligible \cite{mtp15nc,hkt20aom}. In addition, and equally important, the TPA would not affect the band structure of graphene metasurfaces (the frequency of the photonic bands), as it would merely increase the optical loss of the corresponding photonic states. Moreover, the pump power propagating in an optical mode of a PhC with field distribution, $\mathbf{E}_p(\mathbf{r})$, can be expressed as follows \cite{s05book,53}:
\begin{eqnarray}
P_p=\frac{v_g}{4a}\int_{V_{cell}}\frac{\partial}{\partial\omega}\left[\omega\epsilon_{m}(\mathbf{r};\omega)\right]\left|\mathbf{E}_p(\mathbf{r})\right|^2d\mathbf{r},
\label{pow}
\end{eqnarray}
where $v_g=d\omega/dk_x$ is the GV, $V_{cell}$ is the volume of the unit cell, and $\epsilon_{m}(\mathbf{r};\omega)$ is the spatial distribution of the frequency-dependent permittivity of the metasurface (or PhC), that is $\epsilon_{m}(\mathbf{r};\omega)=\epsilon_0$ [$\epsilon_{m}(\mathbf{r};\omega)=\epsilon_g(\omega)$] if $\mathbf{r}$ corresponds to a point in air (graphene). Note that for our graphene structure the waveguide dispersion is much larger than the material dispersion of graphene, so that the latter one can be discarded by setting $\epsilon_{m}(\mathbf{r};\omega)\equiv\epsilon_{m}(\mathbf{r})$.
\begin{figure}[!t]
\centering
\includegraphics[width=\columnwidth]{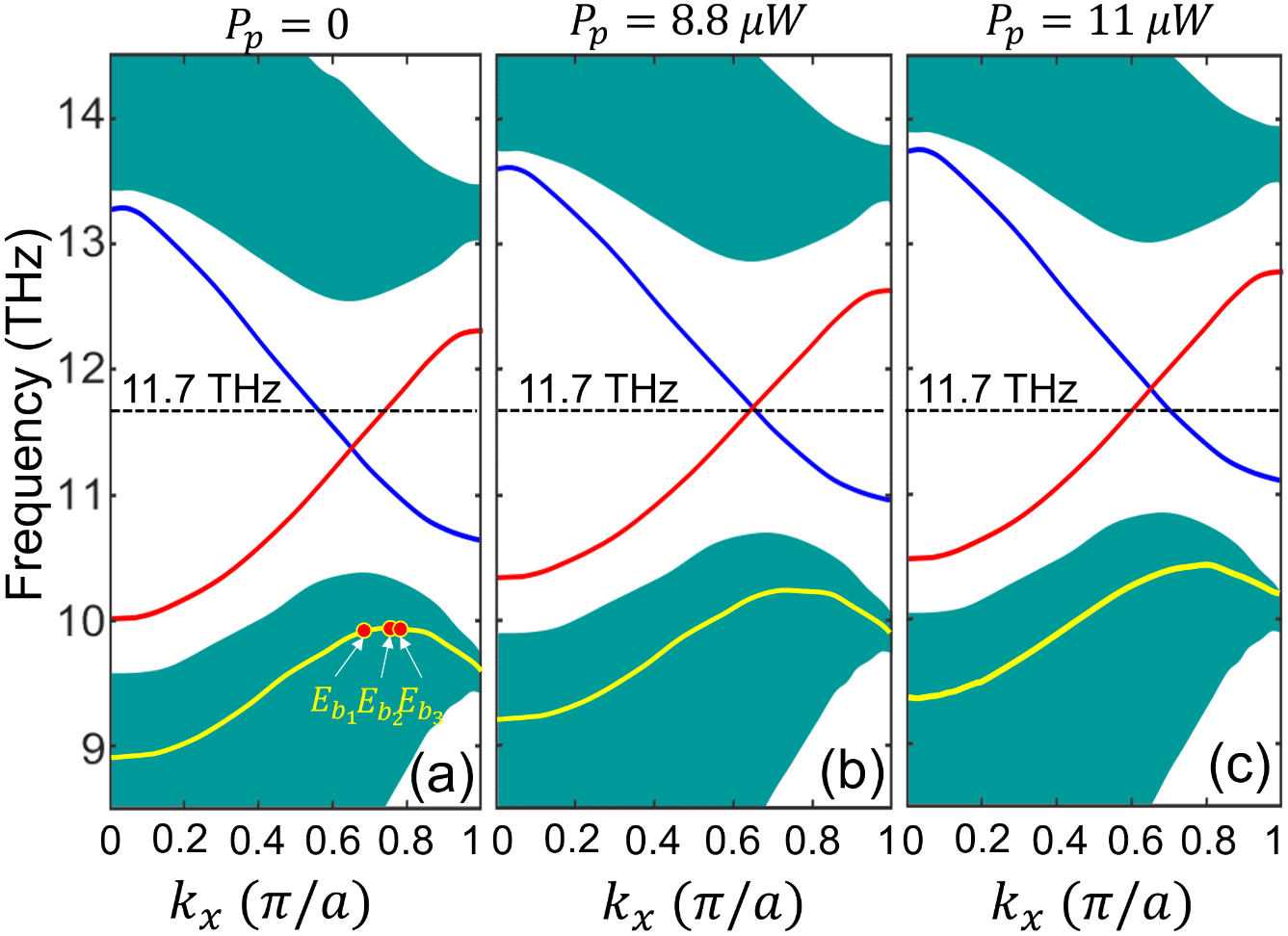}
\caption{From leftmost to the rightmost panels, projected band diagrams of the composite graphene metasurface, calculated for the unperturbed photonic system ($P_{p}=0$), as well as for a photonic system in which the pump power is $P_{p}=\SI{8.8}{\micro\watt}$ and $P_{p}=\SI{11}{\micro\watt}$, respectively. The pump power is inserted in the mode labelled $E_{b_{1}}$. The horizontal dashed line indicates the frequency of an optical signal to be transferred from the edge mode into the topological waveguide mode.}
\label{cap4}
\end{figure}

Since the Kerr coefficient of graphene is particularly large, one can induce a significant variation of the refractive index of the metasurface even at moderate values of the applied local electric field, $\mathbf{E}_p(\mathbf{r})$. This in turn will lead to sizeable frequency shift of the mode dispersion curves. To quantify this effect, we have determined numerically the projected band structure of the composite graphene metasurface and, implicitly, the frequency dispersion curves of the edge and topological waveguide modes. Importantly, in our calculations we have incorporated the contribution of the pump beam to the calculated projected band structure of the composite metasurface. Specifically, we have implemented using COMSOL an algorithm consisting of three main steps \cite{pbo03ol,pbo04josab,pbo04oe}. First, for a given optical pump power and corresponding distribution of the electric permittivity of the metasurface, $\epsilon^{\prime}_{m}(\mathbf{r})$, we determined the field profile corresponding to the bulk mode used as the pump beam. Then, we used \eqref{deln} to compute the local variation of the refraction index, $\Delta n(\mathbf{r})$, and subsequently the updated spatial distribution of the electric permittivity of the metasurface,  $\epsilon^{\prime\prime}_{m}(\mathbf{r})=\epsilon_{0}[n_{0}(\mathbf{r})+\Delta n(\mathbf{r})]^{2}$, with $n_{0}(\mathbf{r})=\sqrt{\epsilon_{m}(\mathbf{r})/\epsilon_{0}}$ being the spatial distribution of the index of refraction of the unperturbed metasurface. In the last step, the projected band structure was determined using the new permittivity distribution, $\epsilon^{\prime\prime}_{m}(\mathbf{r})$. These steps were repeated until the difference $\vert\epsilon^{\prime}_{m}(\mathbf{r})-\epsilon^{\prime\prime}_{m}(\mathbf{r})\vert_{\mathbf{r}\in V_{cell}}$ was smaller than a certain threshold. Note that, due to the very small relative change of the electric permittivity induced by the pump, one iteration sufficed.
\begin{figure}[!t]
\centering
\includegraphics[width=\columnwidth]{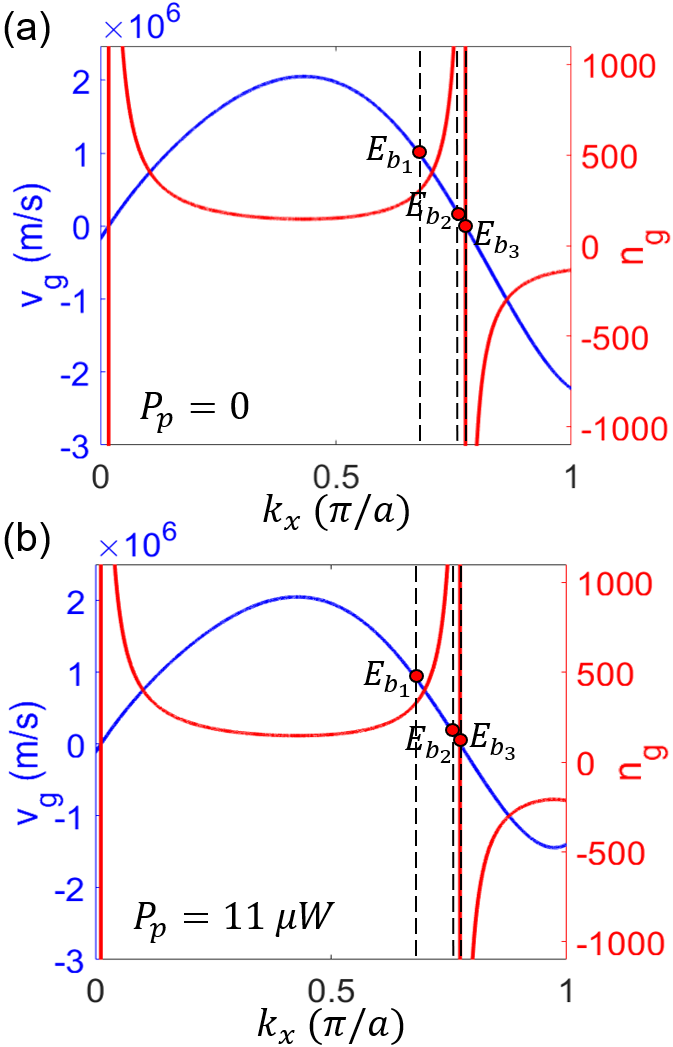}
\caption{(a), (b) Dependence on the wave vector $k_x$ of the group velocity $v_g$ (blue curves) and group index $n_g$ (red curves) of the bulk mode indicated by the yellow curve in figures \ref{cap4}(a) and \ref{cap4}(c), determined for a pump power of $P_{p}=0$ and $P_{p}=\SI{11}{\micro\watt}$, respectively. The pump propagates in the mode labelled $E_{b_{1}}$.}
\label{cap5}
\end{figure}

The results pertaining to the calculation of the projected band structure of the metasurface under the influence of Kerr effect are presented in figure \ref{cap4}. The projected band structures plotted in this figure have been determined for increasing values of the pump power, $P_{p}$. However, in all these calculations we used the same pump mode, \textit{i.e.}, the mode $E_{b_{1}}$ at frequency of \SI{9.9}{\tera\hertz} $(\lambda=\SI{30.3}{\micro\meter})$ in figure \ref{cap2}(b). These results show that, when injecting in the mode $E_{b_{1}}$ pump power with increasing values of \SI{8.8}{\micro\watt} and \SI{11}{\micro\watt}, the projected band structure of the metasurface is blue shifted by about \SI{0.4}{\tera\hertz} and \SI{0.5}{\tera\hertz}, respectively, as compared to the projected band structure of the unperturbed metasurface ($P_{p}=0$).

A consequence of this blue shift of the projected band diagram is that, for a given frequency inside the nontrivial bandgap, the mismatch between the wave-vectors of the topological and edge modes, $\Delta k_x=k_{x,t}-k_{x,e}$, varies with the inserted pump power. Consequently, for a given frequency of the signal, the phase-matching condition defined by $\Delta k_x=0$ can be satisfied by properly tuning the pump power. In other words, the optical coupling between the edge and topological modes can be effectively controlled by tuning the pump power coupled into the bulk mode.

Apart from this blue shift of the projected band diagram, the overall characteristics of the frequency dispersion of the system are practically unchanged. This is demonstrated by the fact that the main dispersion coefficients of the modes depend only weakly on the pump power. This conclusion is illustrated by the results presented in figure \ref{cap5}, where we plot the GV, $v_g$, of the pump mode and its group index, $n_{g}$, defined as $n_{g}=c/v_g$. These physical quantities have been determined for an unperturbed composite metasurface ($P_{p}=0$) and for a metasurface for which $P_{p}=\SI{11}{\micro\watt}$. Note that we chose to focus on the influence of the pump power in the bulk modes on the optical response of the metasurface because, as it is apparent from \eqref{pow}, the pump power employed to optically tune the system depends on the GV of the bulk mode.

In the case when $P_{p}=0$, which corresponds to figure \ref{cap5}(a), the dispersion curve of the GV shows two zero-group-velocity (ZGV) points defined as $v_{g}(k_{x})=0$, at $k_{x}\approx0$ and $k_{x}\approx0.775(\pi/a)$. These ZGV points define two SL regions where the group index has particularly large values. Specifically, as $k_{x}$ is tuned from the bulk mode $E_{b_{1}}$ [$k_{x}=0.68(\pi/a)$] to the bulk mode $E_{b_{3}}$ [$k_{x}=0.774(\pi/a)$], $v_{g}$ decreases and approaches zero, resulting in an increase of the group index to over 1000. This physical picture changes only slightly when $P_{p}\neq0$. The main difference in the case when a pump power is applied is that the values of the optical dispersion coefficients change with the pump power, especially near the edge of the first Brillouin zone, namely near $k_{x}=\pi/a$.

The property of extremely small group velocity of the bulk modes can be employed to optimize the operation of the nonlinear optical coupler. To be more specific, it can be seen from \eqref{pow} that for a given pump power $P_{p}$, the smaller the GV of the mode (that is, the deeper in the SL regime the device is operated), the larger an optical field is generated and consequently a larger variation of the index of refraction of the metasurface is achieved. Therefore, for a given operating pump power, the smaller the GV is, the larger is the induced frequency shift of the projected band structure. As a design principle, it becomes apparent that in order to attain an optimum operation of our nonlinear optical coupler, namely to achieve a minimum operating pump power, the system parameters must be chosen in such a way that the phase-matching condition $\Delta k_x=0$ is fulfilled in the SL region.

\subsection{Analysis of the phase-matching condition}
\begin{figure}[!t]
\centering
\includegraphics[width=\columnwidth]{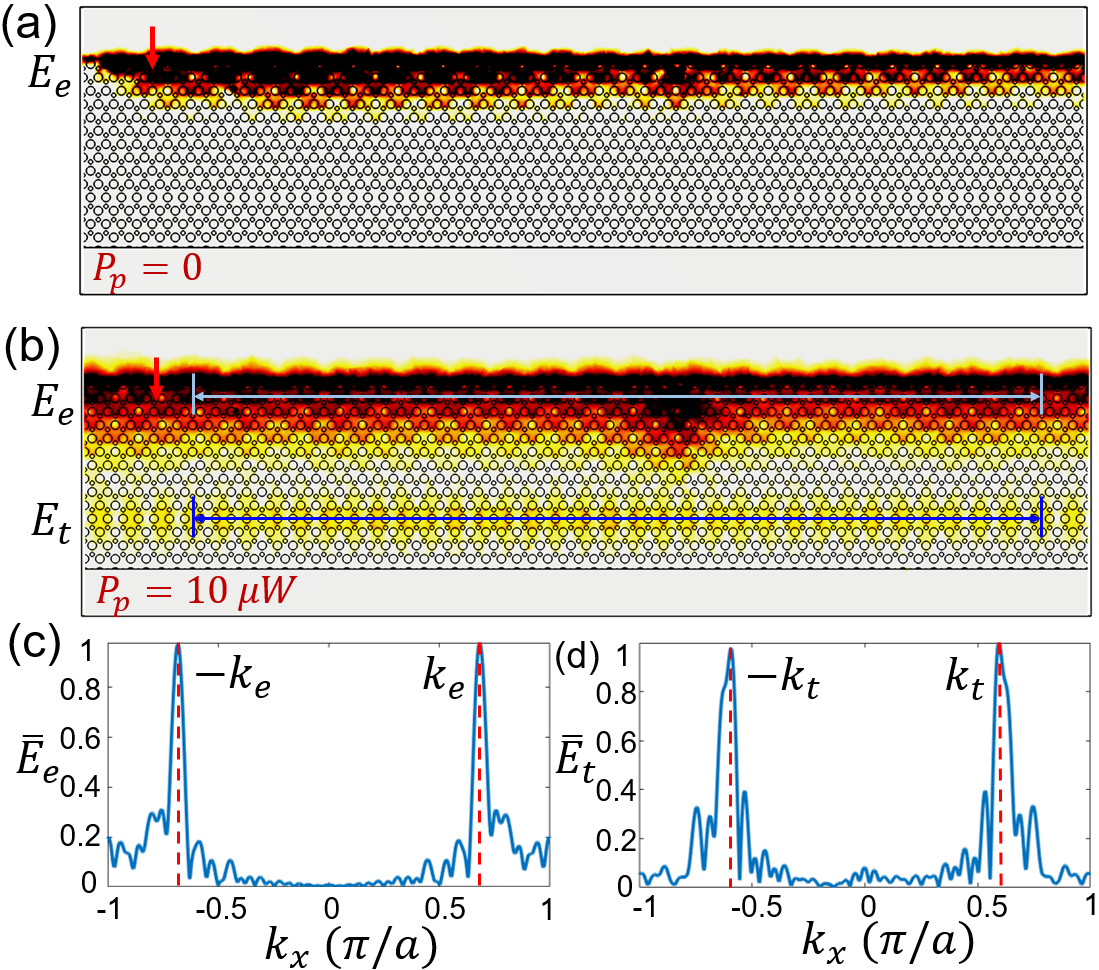}
\caption{(a), (b) Spatial field profile in the nonlinear coupler, determined for $P_{p}=0$ and $P_{p}=\SI{10}{\micro\watt}$, respectively. The input pump power is injected in the bulk mode $E_{b_{1}}$ at frequency of $\SI{11.7}{\tera\hertz}$, \textit{via} a dipole source indicated by the red arrow. (c), (d) Fourier transform of normalized electric fields of the trivial edge mode, $\bar{E_e}$, and topological mode, $\bar{E_t}$, shown in figure \ref{cap6}(b), respectively. The spatial intervals of the Fourier-transform calculations of the edge and topological modes are marked by light and dark blue lines in figure~\ref{cap6}(b), respectively.}
\label{cap6}
\end{figure}
Since the difference between the wave-vectors of the topological and edge modes is the key physical parameter that governs the efficiency of the optical coupling between the two modes, in what follows we focus our attention on the dependence of this physical quantity on the pump power, $P_{p}$.

To begin with, we assume that the pump power is inserted in the graphene metasurface using the bulk mode $E_{b_{1}}$. For this configuration, using full-wave numerical simulations based on the finite-element method, we determined the spatial field profile in the nonlinear optical coupler for pump power values of $P_{p}=0$ (unperturbed metasurface) and $P_{p}=\SI{10}{\micro\watt}$, the corresponding results being presented in figures \ref{cap6}(a) and \ref{cap6}(b), respectively.

We used absorbing boundary conditions, and placed the boundaries of the computational domain in such a way that they are slightly separated from the graphene metasurface by a thin layer of air. By avoiding the graphene metasurface to reach the boundaries of the computational domain one improves the accuracy and stability of the simulations. Moreover, the transverse width of the upper domain of the graphene structure has width of $10$ unit cells and was chosen to be larger than that of the lower domain, which has width of $3$ unit cells, because both the overlap between the optical fields of the two modes and the field distribution of the bulk pump mode are primarily localized in the upper domain of the graphene metasurface. In addition, with this choice the simulation time decreases. Finally, the edge mode is excited \textit{via} a dipole source (red arrow) with frequency of $\SI{11.7}{\tera\hertz}$, placed in the vicinity of the top boundary of the graphene metasurface with length of $65$ unit cells along the longitudinal direction.

Our calculations show that when $P_{p}=0$, the wave-vector mismatch is relatively large, $\Delta k_x=0.28\pi/a$. As expected, because of this significant wave-vector mismatch only a small amount of optical power couples into the topological mode, a fact illustrated by figure \ref{cap6}(a), too. By contrast, when the pump power injected in the bulk mode $E_{b_{1}}$ increases to $P_{p}=\SI{10}{\micro\watt}$, the spatial field distribution presented in figure \ref{cap6}(b) reveals an efficient optical coupling between the topological and edge modes. This conclusion is also supported by the significantly reduced value of the wave-vector mismatch, which in this case is $\Delta k_x=0.032\pi/a$. This demonstrates that the amount of optical power transferred from the edge mode to the topological waveguide mode can be effectively tuned by optically tuning the wave-vector mismatch between the two modes \textit{via} the pump power $P_{p}$ propagating in the bulk mode.

The wave-vector mismatch can also be extracted from the spatial field profiles presented in figures \ref{cap6}(a) and \ref{cap6}(b), which provides an alternative approach to quantitatively analyze the operation of the proposed nonlinear optical coupler. This approach consists in Fourier transforming the field profiles, $\mathbf{E}(\mathbf{r})$, to the momentum space, $\bar{\mathbf{E}}(\mathbf{k})$, and subsequently extracting the value of the wave-vectors corresponding to the peaks of the field spectra. This procedure is illustrated by the spectra plotted in figures \ref{cap6}(c) and \ref{cap6}(d), which correspond to the edge and topological modes observed in figure \ref{cap6}(b), respectively. These spectra, corresponding to a pumped metasurface, display two symmetrically located peaks pertaining to forward- $(k_{e}, k_{t})$ and backward-propagating $(-k_{e}, -k_{t})$ modes.

To gain deeper insights into the physics of the optical coupling between the edge and topological modes, we developed a coupled-mode theory (CMT) that describes the dynamics of the mode amplitudes upon their propagation in the graphene metasurface. To this end, consider the $x$-dependent amplitudes of the edge and topological mode, $E_e(x)$ and $E_t(x)$, respectively, in a case characterized by the wave-vector mismatch, $\Delta k_x$. Then, their dependence on the propagation distance, $x$, is governed by the following system of coupled ordinary differential equations \cite{54}:
\numparts
\begin{eqnarray}
\label{couplfa}&\frac{dE_e(x)}{dx}=i\kappa E_t(x)e^{-i\Delta k_xx}-\frac{\alpha}{2}E_e(x),\\
\label{couplfb}&\frac{dE_t(x)}{dx}=i\kappa E_e(x)e^{i\Delta k_xx}-\frac{\alpha}{2}E_t(x),
\end{eqnarray}
\endnumparts
where $\alpha$ is the loss coefficient due to intrinsic and radiative losses and $\kappa$ is the coupling coefficient between the modes. Note that, for the sake of simplicity, we assumed that the loss coefficients of the two modes are equal.

After simple mathematical manipulations, one can derive from \eqref{couplfa} and \eqref{couplfb} the dependence on the propagation distance of the physical quantities of interest, namely the mode powers $P_e(x)=\vert E_e(x)\vert^2$ and $P_t(x)=\vert E_t(x)\vert^2$. Subject to the initial conditions $E_e(0)=E_0$ and $E_t(0)=0$, the optical power of edge and topological interface modes can be expressed as:
\numparts
\begin{eqnarray}
\label{couplpa}&P_e(x)=\frac{P_0}{\xi^2}e^{-\alpha x}\left[\xi^2\cos^2(\xi\kappa x)+\rho^{2}\sin^2(\xi\kappa x)\right],\\
\label{couplpb}&P_t(x)=\frac{P_0}{\xi^2}e^{-\alpha x}\sin^2(\xi\kappa x),
\end{eqnarray}
\endnumparts
where $P_0=\vert E_0\vert^2$ is the initial power inserted in the edge mode, $\rho=\Delta k_x/(2\kappa)$ is a parameter that measures the relative coupling strength, and $\xi=\sqrt{1+\rho^{2}}$ is a wave-vector scaling factor. It can be easily verified using \eqref{couplpa} and \eqref{couplpb} that the total power in the two modes satisfies the power decay relation, $P_e(x)+P_t(x)=P_0e^{-\alpha x}$.
\begin{figure}[!b]
\centering
\includegraphics[width=\columnwidth]{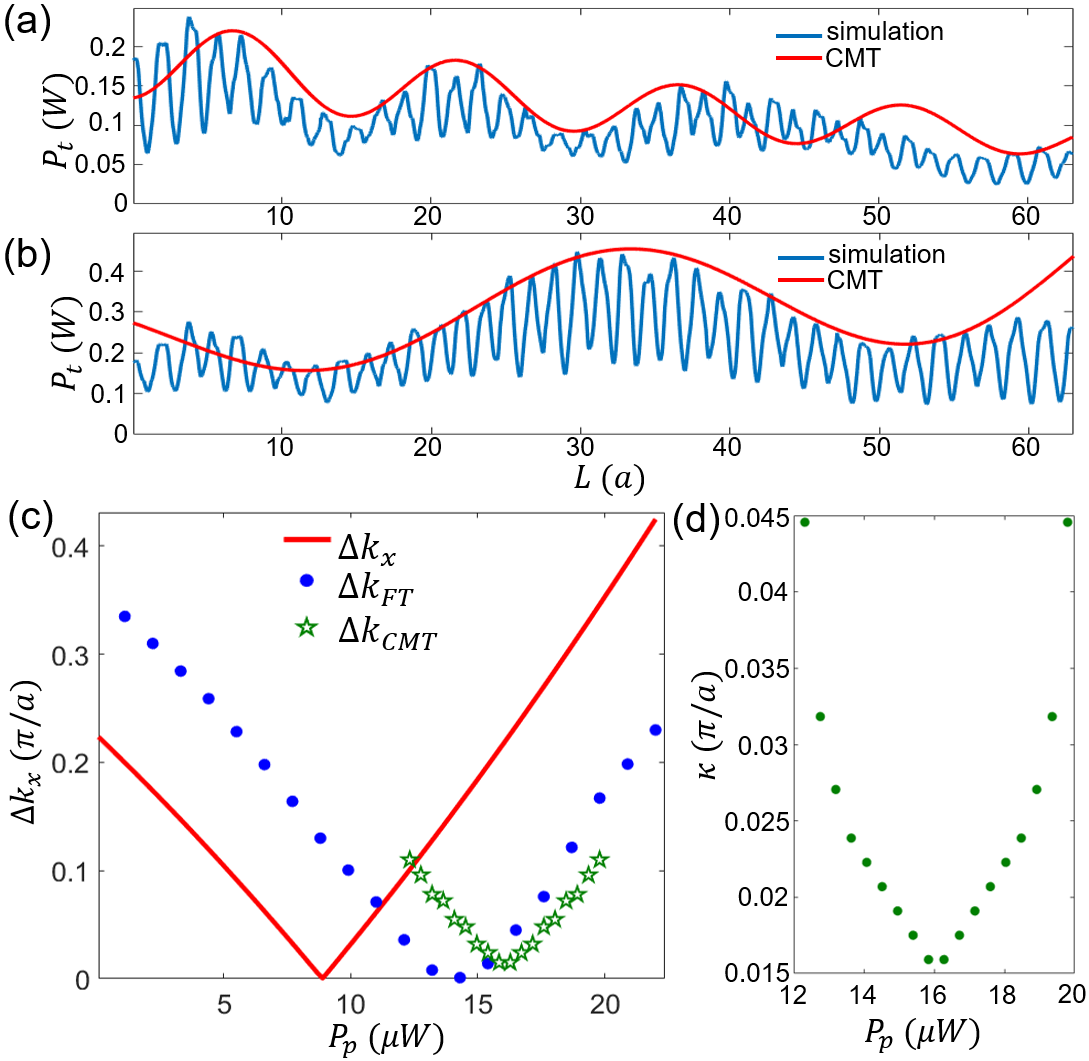}
\caption{(a), (b) Power of the topological interfacial mode, $P_t$, vs. the propagation length $L$, determined in the regimes of weak coupling ($P_{p}=\SI{12.3}{\micro\watt}$) and optimum coupling $(P_{p}=\SI{15.4}{\micro\watt}$), respectively, as predicted by the CMT. Blue and red curves correspond to full-wave simulations and fitted solutions of the CMT, respectively. (c) Dependence of wave-vector mismatch, $\Delta k_x$, on the pump power $P_{p}$ injected in the bulk mode $E_{b_{1}}$. The wave-vector $\Delta k_x$ is extracted from the projected band diagram (red line), Fourier transform of the spatial distribution of the propagating fields (dots), and CMT (stars). (d) Coupling coefficient $\kappa$ \textit{vs}. pump power $P_{p}$, predicted by the CMT.}
\label{cap7}
\end{figure}

The CMT provides us an alternative approach to determine the wave-vector mismatch between the edge and topological modes, and consequently the means to validate the physical assumptions on which the CMT is based. In particular, by fitting the powers $P_e(x)$ and $P_t(x)$ determined using full-wave simulations with the analytic solutions \eqref{couplpa} and \eqref{couplpb}, one can extract the parameters $\kappa$ and $\Delta k_x$ of the nonlinear optical coupler. This procedure and the corresponding results are summarized in figure \ref{cap7}. Thus, we present  in figures \ref{cap7}(a) and \ref{cap7}(b) the dependence on the propagation distance of the optical power in the topological mode, $P_t(x)$, determined in the regime of weak coupling ($P_{p}=\SI{12.3}{\micro\watt}$) and optimum coupling $(P_{p}=\SI{15.4}{\micro\watt}$), respectively, when the pump power is inserted in the bulk mode $E_{b_{1}}$. In the full-wave simulations the power $P_t$ is computed by integrating the longitudinal component of the Poynting vector, $S_{x}$, across the transverse extent of the interface region.

The top two panels of figure \ref{cap7} show that the CMT solution of \eqref{couplfa} and \eqref{couplfb} provides a good fit for the variation of the envelope of the full-wave solution obtained numerically, suggesting that it properly captures the main physics of the optical structure. The full-wave simulations, on the other hand, describe the power dynamics at the lattice constant scale. Moreover, it can be seen in figure \ref{cap7}(a) that if the pump power $P_{p}=\SI{12.3}{\micro\watt}$, the power $P_t$ determined using the CMT oscillates with period of about $15a$, the wave-vector mismatch and coupling constant being $\Delta k_x=0.11\pi/a$ and $\kappa=0.045\pi/a$, respectively. When the pump power is increased to $P_{p}=\SI{15.4}{\micro\watt}$, which corresponds to the data in figure \ref{cap7}(b), the period of the power oscillations increases to $40a$, whereas $\Delta k_x=0.024\pi/a$ and $\kappa=0.006\pi/a$. It can be said that an optimum coupling regime has been reached as the wave-vector mismatch between the two modes has decreased to a vanishingly small value.

To quantify more completely the influence of the pump power on the efficiency of the mode coupling, we determined the wave-vector mismatch for different values of the pump power using the three methods we just described. The results of these calculations, presented in figure \ref{cap7}(c), demonstrate that the wave-vector mismatch determined using the projected band diagram ($\Delta k_x$), the Fourier transform of the spatial distribution of the field ($\Delta k_{FT}$), and the CMT ($\Delta k_{CMT}$) depends on the pump power in a similar way, the only difference being a slight shift of a few \si{\micro\watt}'s among the corresponding curves. Specifically, in all three cases, the wave-vector mismatch vanishes for a certain value of the pump power $P_{p}$ injected in the bulk mode $E_{b_{1}}$ approaching a certain value, so that all three methods predict that efficient mode matching can be achieved by tuning $P_{p}$.

The CMT allows one to extract from the power dynamics not only the wave-vector mismatch but also the coupling constant $\kappa$. The relevant data, plotted in figure \ref{cap7}(d), show that $\kappa$ reaches a minimum value for the same pump power for which $\Delta k_{CMT}\approx0$, that is when the two modes are phase-matched.

\subsection{Optically controllable mode coupling in the slow-light regime}
As we have previously alluded to, SL effects can potentially be used to reduce the operating pump power of the nonlinear optical coupler. To quantitatively analyze these nonlinear optical effects, we tuned the frequency of the pump beam in such a way that the corresponding GV of the bulk pump mode is significantly reduced and repeated the analysis performed in the case when the pump power is inserted in the bulk mode $E_{b_{1}}$. In what follows, we discuss the conclusions of these computational investigations.
\begin{figure}[!b]
\centering
\includegraphics[width=\columnwidth]{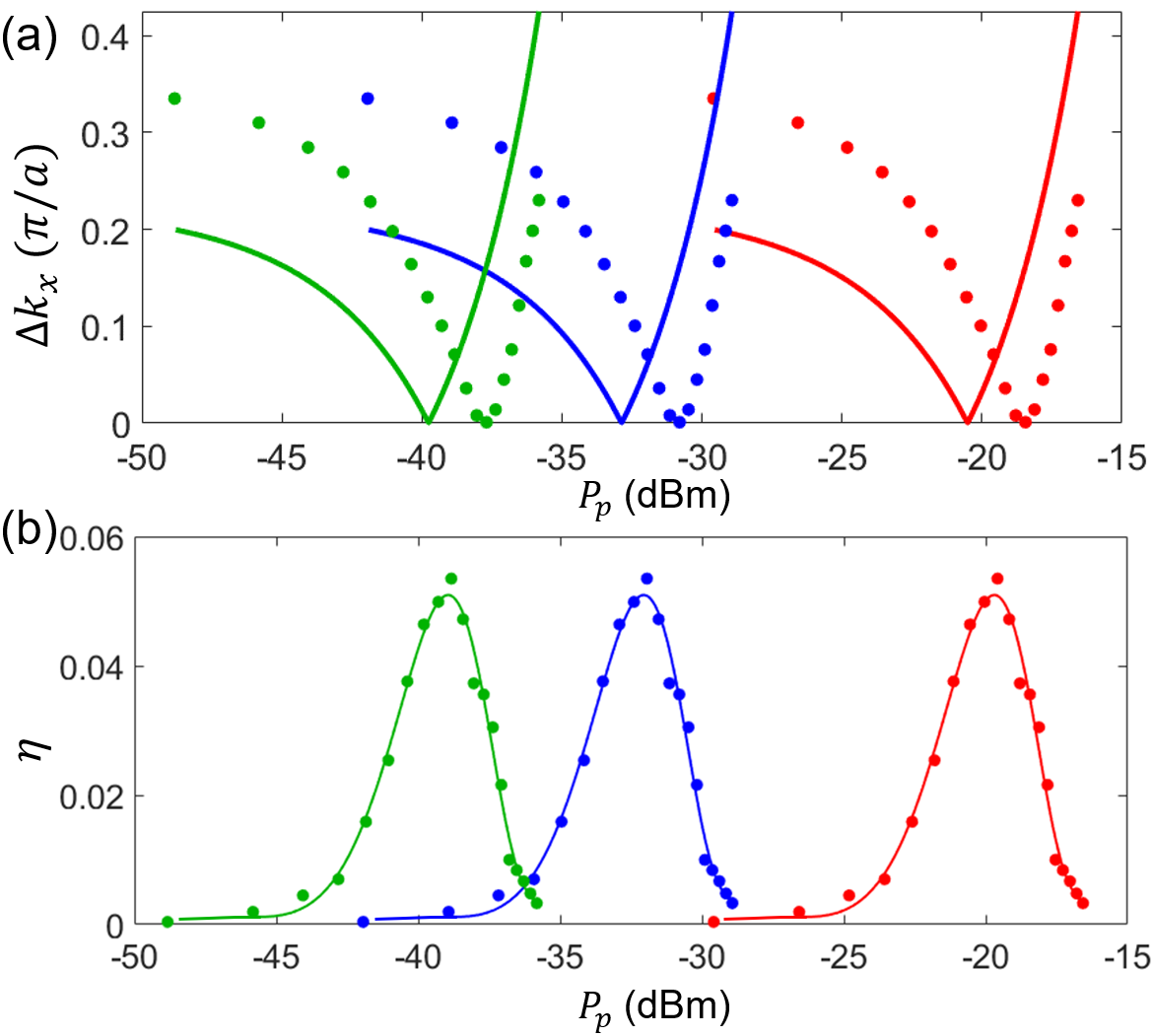}
\caption{(a) Wave-vector mismatch $\Delta k_x$ \textit{vs}. pump power $P_{p}$, when the pump beam propagates in mode $E_{b_{1}}$ (red), $E_{b_{2}}$ (blue), and $E_{b_{3}}$ (green). The solid and dotted curves correspond to $\Delta k_x$ extracted from the projected band diagram and calculated by Fourier transforming the spatial distribution of the field, respectively. (b) Ratio $\eta$ between the output power in the topological mode and the input power in the edge mode \textit{vs}. the pump power, determined for the pump modes $E_{b_{1}}$ (red), $E_{b_{2}}$ (blue), and $E_{b_{3}}$ (green). The solid curves are a guide to the eye.}
\label{cap8}
\end{figure}

To vary the GV of the pump beam we chose its frequency in such a way that the bulk pump mode is tuned from $E_{b_{1}}$ to $E_{b_{2}}$ and finally to $E_{b_{3}}$. In the first part of our analysis, in all these three cases we used the Fourier transform based approach to compute the pump power dependence of the wave-vector mismatch, $\Delta k_x$, the outcome of these calculations being depicted with starred symbols in figure \ref{cap8}(a). In addition to this analysis based on the Fourier transform of the spatial field profiles, we also used the pump power dependent projected band diagrams to extract the value of $\Delta k_x$ and summarize the results of these calculations in figure \ref{cap8}(a) using solid curves.

In all three cases the two methods agree in their prediction of the pump power at which the two modes are phase-matched ($\Delta k=0$). When the phase-matching condition is fulfilled by tuning the pump power to a certain value, the topological and edge modes are most efficiently coupled, meaning that a maximum amount of power can be transferred the two modes. Note that this critical pump power value depends strongly on the GV of the pump beam. In particular, this pump power decreases by more than $100\times$ when the pump mode is pushed deep into the SL domain, namely when it is tuned from mode $E_{b_{1}}$ (red curve) to $E_{b_{3}}$ (green curve).

The most important parameter that characterizes our nonlinear optical coupler is the transmission coefficient, $\eta$, defined as the ratio between the output power in the topological mode and the input power of the edge mode, as it quantifies the efficiency of the optical coupler. Therefore, in the second part of our analysis of the nonlinear optical coupler we have determined the dependence of $\eta$ on the pump power. These computations, whose conclusions are presented in figure \ref{cap8}(b), have been performed for the pump modes $E_{b_{1}}$ (red), $E_{b_{2}}$ (blue), and $E_{b_{3}}$ (green). The pump power dependence of the transmission $\eta$ shows that, as expected, in all cases there is a certain value of the pump power for which maximum transmission is achieved. This pump power at which maximum transmission corresponds to $\Delta k=0$ [cf. figures \ref{cap8}(a) and \ref{cap8}(b)] which demonstrates that the optical power is transferred between the two modes through evanescent mode coupling. Importantly, the pump power required to achieve maximum power transfer decreases as the GV of the pump mode, $v_g$, decreases; however, it is noteworthy to point out that the maximum transferred power does not depend on $v_g$.

\section{\label{sec5}Conclusion}
In conclusion, we have investigated an optically controllable nonlinear coupler that can be used to transfer optical power  between a trivial edge mode and a topological interface mode of a structured graphene metasurface. The nonlinear optical coupler is controlled \textit{via} the Kerr effect in graphene induced by a pump beam. Specifically, the pump beam is employed to tune the band structure of the photonic system and, consequently, the key parameter that defines the efficiency of the nonlinear optical coupler, namely the wave-vector mismatch between the edge and topological interface modes. Importantly, we have also demonstrated that the required pump power can be significantly reduced if the optical device is operated in the slow-light regime. We performed our analysis using both \textit{ab initio} full-wave simulations and a coupled-mode theory that captures the main physics of this active coupler and observed a good agreement between the two approaches. Practical technological implications of this work have been discussed, too.

\section*{Data availability statement}
The data that support the findings of this study are available upon reasonable request from the authors.

\section*{Funding}
China Scholarship Council; UCL.

\section*{Conflict of interest}
The authors declare no conflicts of interest.

\section*{Acknowledgments}
We thank Dr. Jian Wei You for numerous illuminating discussions on computational aspects of this study.

\end{document}